\documentstyle[a4,12pt]{article}

\def\vec#1{\ifmmode
\mathchoice{\mbox{\boldmath$\displaystyle\bf#1$}}
{\mbox{\boldmath$\textstyle\bf#1$}}
{\mbox{\boldmath$\scriptstyle\bf#1$}}
{\mbox{\boldmath$\scriptscriptstyle\bf#1$}}\else
{\mbox{\boldmath$\bf#1$}}\fi}

\begin{document}

\markboth{Authors' Names}
{Instructions for Typing Manuscripts (Paper's Title)}


\title{Quantum dynamics and Random Matrix theory}

\author{\footnotesize Herv\'{e} KUNZ\footnote{hkunz@dpmail.epfl.ch}\\
\small{Institute of theoretical Physics, Swiss Federal Institute of
Technology Lausanne} \\ \small{Lausanne , CH-1015 EPFL,
Switzerland} }

\maketitle

\begin{abstract}
We compute the survival probability of an initial state, with an energy in a
certain window, by means of random matrix theory. We determine its probability
distribution and show that is is universal, i.e. caracterised only by the
symmetry class of the hamiltonian and independent of the initial state.
\end{abstract}

\vspace{1cm}

In classical mechanics, temporal chaos is caracterised by the extreme
sensibility of a trajectory to variation of initial conditions. No direct analog
of this phenomenon has been found in quantum mechanics so far. On the other
hand, numerical evidence has been accumulated [1], showing that energy
levels of a quantum system, whose classical counterpart is chaotic, have a
statistical behavior described by Wigner's random matrix theory (RMT), on the
mean level spacing scale.
\noindent The question we want to adress is the following: are there specific
predictions of RMT for quantum dynamics, which would caracterise the temporal
behavior of "chaotic" quantum systems.

\noindent We consider the following situation: The system is prepared in an
initial state $\varphi$ at time $0$, with an energy in a certain window,
centered at $e$ and of width $2sl(e)$, where $l(e)$ is the mean level spacing,
and we want to compute the probability to find our system again in the state
$\varphi$, at a later time $t$. This quantity that we call the \emph{survival
probability} $R$ is given by
\begin{equation} \label{survivalprobability}
R = \left | \frac{\left ( \varphi | \exp{i \frac{t}{\hbar} H}\,\,
P(\Delta) \varphi \right)}{\left ( \varphi, \, P(\Delta) \varphi \right)} \right
|^2
\end{equation}
\noindent $H$ is the hamiltonian of our system and $P(\Delta)$ is the spectral
projector on $\Delta$.

\noindent We have chosen to take an energy in a range of the order of the mean
level spacing in order to look at properties of the system which are independent
of specific details.
If $\left ( \lambda_j, \, \psi_j  \right)$ denote respectively the $j^th$
eigenvalue and eigenvector of the Hamiltonian $H$, then the survival probability
can be written as
\begin{equation} \label{2}
R = \left | \frac{ \sum_{j=1} \, y_j \chi_j \exp{2 \pi i \tau
x_j}}{\sum_{j=1} \, y_j \chi_j} \right | ^2 \, \Theta \left ( \sum_j \, \chi_j
-1 \right )
\end{equation}
where
\begin{equation} \label{3}
y_j = \left | \left ( \varphi, \, \psi_j \right ) \right |^2
\end{equation}
and if we define $x_j$ by the relation
\begin{equation} \label{4}
\lambda_j = e + x_j l(e)
\end{equation}
\begin{equation} \label{5}
\chi_j \equiv \chi_{(-s, s)} (x_j) = \left \{ \begin{array}{l} 1 \quad
\textrm{if} \, |x_j| \leq s \\ 0 \quad \textrm{otherwise} \end{array} \right.
\end{equation}

The Heaviside function $\Theta$ ensures that there is at least one eigenvalue in
$\Delta$. What appears naturally in this expression is the time measured in
units of the Heisenberg time
\begin{equation} \label{6} t_H = \frac{h}{l(e)} \end{equation}
so that \begin{equation} \label{7} \tau = \frac{t}{t_H}. \end{equation}

\noindent If we look at this problem from the point of view of RMT, we will
replace the Hamiltonian by a large $N \times N$ self-adjoint matrix, whose
probability distribution is basis independent and therefore of the form
\begin{equation} \label{8}
e^{- W \left ( \lambda_1, \ldots, \lambda_N \right)} \, dH
\end{equation}
Wigner's gaussian model corresponds to the choice
\begin{equation} \label {9}
W = \frac{N}{2} \, \sum_{j=1}^{N} \, \lambda_j^2
\end{equation}

\noindent The first conclusion to be drawn is that the survival probability is
statistically \emph{independent of the initial state $\varphi$}. This follows
from the fact that the variables $\left \{ y_j \right \}_{j=1}^{N}$ have a
probability distribution, independent of $\varphi$ and given by:
\begin{equation} \label{10}
\mu_N(\vec{y}) d\vec{y} = \frac{1}{C_N} \, \delta \left ( \sum_{j=1}^N \, y_j
-1 \right) \, \prod_{j=1}^N \, y_j^{\frac{\beta}{2}-1} \, d\vec{y}
\end{equation}

The parameter $\beta=1, \, 2, \, 4$ caracterise the symmetry class of the
Hamiltonian, respectively orthogonal, unitary and symplectic. Equation
(\ref{10}) follows easily from the Haar measure on the corresponding groups.
$C_N$ is a normalising constant.

The variables $\left \{ x_j \right \}_{j=1}^N$ are statistically independent of
the variables $\left \{ y_j \right \}_{j=1}^N$ and have a distribution given by
\begin{equation} \label{11}
\frac{1}{D_N} \exp{-W\left ( e + \vec{x} l(e) \right)} \Delta^{\beta} \left (
\vec{x} \right) d\vec{x}
\end{equation}
where the Van der Monde determinant
\begin{equation} \label{12}
\Delta \left ( \vec{x} \right) = \prod_{1 \leq i < j \leq N}\, \left | x_i - x_j
\right |
\end{equation}
comes from the change of variables $H_{ij} \rightarrow \left ( \lambda_j, \,
\psi_j \right )_{j=1}^N$ [2]. $D_N$ is a constant of normalisation.

We can take $l(e)= \frac{1}{N \rho(e)}$, where $\rho(e)$ is the density of
states when $N=\infty$.

\noindent The problem that we need to solve now is to find the probability
distribution of the survival probability $p(R) dR$ in the $N=\infty$ limit.
We find that $R$ is \textbf{not} \emph{self-averaging} i.e. $p(R)$ is
not a delta distribution concentrated on the mean value of $R$. On the
other hand its probability distribution $p(R)$ is \emph{universal},
i.e. it depends only on the symmetry parameter $\beta$, at least for a large
class of $W$.

\noindent There are two formulas for $p(R)$, one more appropriate to small
windows, another one to large windows.

\noindent In the first case, we decompose $p(R)$ into
\begin{equation} \label{13}
p\left ( R | \tau \right) = \sum_{n=1}^{\infty} \, \frac{E_n}{1-E_0}
 p_n \left ( R | \tau \right)
 \end{equation}
 where $E_n$ is the probability to find exactly $n$ eigenvalues in $\Delta$ and
$p_n(R|\tau)$ is the conditional probability density of $R$
knowing that there are exactly $n$ eigenvalues in $\Delta$.

\noindent It can be expressed as
\begin{equation} \label{14}
p_n \left ( R | \tau \right) = \int_{-s}^{s} \, \hat{E}\left ( x_1,
\ldots, x_n \right) d^n x \int_{0}^{\infty} \, \mu_n \left (z_1, \ldots, z_n
\right) d^n z \delta \left ( R - \left | \sum_{j=1}^{n} \, z_j \exp{2\pi i \tau
x_j} \right |^2 \right)
\end{equation}
\begin{equation} \label{15} \hat{E} \left (x_1, \ldots , x_n \right ) =
\frac{E \left (x_1, \ldots , x_n \right )}{E_n}
\end{equation}
\begin{equation} \label{15a} E_n = \int_{-s}^{s} \, E\left (x_1, \ldots , x_n
\right ) d^n x
\end{equation}

\noindent $E   \left (x_1, \ldots , x_n \right )$ being the probability density
of finding the $n$ eigenvalues in $\Delta$ at $ \left (x_1, \ldots , x_n \right
)$.

\noindent Useful expressions for $E  \left (x_1, \ldots , x_n \right )$ and
$E_n$ can be found in [2] and [3]. It is expressible in terms of a determinant
\begin{equation} \label{16}
E  \left (x_1, \ldots , x_n \right ) = \textrm{det} \, L_{\beta} \left ( x_i |
x_j \right) \quad ; \quad (i,j) \in (1 \ldots n)
\end{equation}
where
\begin{equation} \label{17}
L_\beta = \frac{K_\beta}{1 - K_\beta}
\end{equation}
$K_\beta$ is an operator whose kernel in the simplest case $\beta=2$ is given by
\begin{equation} \label{18}
K_\beta \left ( x |y \right) = \frac{\sin{\pi (x-y)}}{\pi (x-y)}
\end{equation}
defined on $L^2(-s, \,s )$.

Universality comes from the fact that $E \left ( x_1, \ldots, x_n \right)$ is
expressible in terms of the correlation functions and the latter ones depends
only on $\beta$, for a large class of $W$. $W$ modifies only the density of
states and therefore the mean level spacing $l(e)$.

\noindent This expression for $p \left ( R | \tau \right)$ is mostly useful in
the small window limit, because when $s \rightarrow 0$
\begin{equation} \label{19}
E_n \sim s^{\frac{\beta}{2} n^2 + n(1-\frac{\beta}{2})}
\end{equation}
Moreover in this case we have
\begin{equation} \label{20}
\lim_{s \rightarrow 0} s^n \hat{E} \left ( s x_1, \ldots, s x_n \right) = A_n
\prod_{1 \leq i < j \leq n}\, \left | x_i-x_j \right |^\beta
\end{equation}
so that the probability distribution of $R$ shows a \emph{scaling
behavior}
\begin{equation} \label{21}
\lim_{\begin{array}{c}s \rightarrow 0 \\ R \rightarrow 1 \end{array}}
s^{-\beta -1}pr \left \{ \frac{1 - R}{\left ( \pi \tau s \right ) ^2}
\geq x \right \} = \int_{x}^{1} \, g_{\beta} (\lambda) d\lambda
\end{equation}
the function $g_\beta (\lambda)$ being given by
\begin{equation} \label{22}
g_\beta (\lambda) = A_\beta \lambda^{\frac{\beta -1}{2}} \left [ \frac{1}{2}
\sqrt{1 - \lambda} + \frac{1}{2} \ln{\lambda} - \ln{1 + \sqrt{1-\lambda}} \right
] \end{equation}

On the other hand, one can see from eq (\ref{13}) and (\ref{14}) that the
probability distribution of $R$ is well defined at \emph{infinite
times}. Namely
\begin{equation} \label{23}
p_n \left ( R | \tau \right) = p_n \left ( R | \infty
\right) + O \left ( \frac{1}{\tau} \right)
\end{equation}
as can be seen by an integration by parts where
\begin{equation} \label{24}
p_n \left (R | \infty  \right ) = \int_{0}^{\infty}\, \mu_n \left (
z_1, \ldots , z_n \right) \int_{0}^{2 \pi} \, \prod_{j=1}^{n} \,
\frac{d\phi_j}{2 \pi} \delta \left ( R - \left | \sum_{j=1}^{n}\, z_j
e^{i \phi_j} \right | ^2 \right )
\end{equation}
\noindent Using an integral representation for the delta appearing in the definition
(\ref{10}) of the $\mu_n$, we can reexpress (\ref{24}) as
\begin{equation} \label{25}
p_n \left ( R | \infty \right) = \frac{1}{4 \pi c_n}
\int_{\epsilon -i \infty}^{\epsilon + i \infty} du \,e^u \int_{0}^{+ \infty}
dr \, r J_0 \left ( \sqrt{R} r \right ) \left [ \int_{0}^{\infty} dz
\, e^{-uz} J_0 (rz) \, z^{\frac{\beta}{2}-1} \right ]^n
\end{equation}

\noindent $\epsilon$ being any positive number, and $J_0(x)$ the Bessel function.
This expression can be simplified, considerably when $\beta = 2, \, 4$.

\noindent In the unitary case $(\beta = 2)$ one finds
\begin{equation} \label{26}
p_n\left (R | \infty \right ) = \frac{n -1}{2} \left ( 1 - R
\right)^{\frac{n-3}{2}}
\end{equation}

\noindent For a \emph{large window} of energy, it is more appreciate to find
another expression for $p_n \left (R | \tau \right )$. It is given as some
integral over a Fredholm determinant  $\mathcal{G}$.

\noindent $\mathcal{G}$ is a generating function for the variables $\left \{ y_j
\right \}$ and $\left \{ x_j \right \}$ appearing in the definition of $R$, eq
(\ref{2}).
\begin{equation} \label{27}
\mathcal{G} \left (r ; \varphi; z \right ) = \lim_{N \rightarrow \infty} \left
< \exp{-i N \sum_{j=1}^{N} \, y_j \chi_j \left [ r \cos{\left (2 \pi \tau x_j +
\varphi \right )} + z \right ]} \right >
\end{equation}

\noindent It can be expressed in terms of the operator $K_\beta$ appearing in eq
(\ref{18}), when $\beta= 1,\, 2$ as
\begin{equation} \label{28}
\mathcal{G} = E_0 \left [ \textrm{det}\, \left ( 1 + K_\beta g^{\frac{\beta}{2}}
\right ) \right ]^{\frac{\beta}{2}}
\end{equation}
with
\begin{equation} \label{29}
E_0 = \left [ \textrm{det} \, \left ( 1 - K_\beta \right)
\right]^{\frac{\beta}{2}}
\end{equation}
and $g$ is the multiplication operator by the function
\begin{equation} \label{30}
g = \left [ 1+ \frac{2i}{\beta} \left [ z + r \cos{(2 \pi \tau x + \varphi)}
\right ] \right ]^{-1}
\end{equation}

When the window is large $(s>>1)$ we can expand the determinant in
powers of $K_\beta$, the first two terms of this expansion dominating the other
ones [4].

\noindent One finds that the probability distribution is \emph{exponential}.
\begin{equation} \label{31}
\lim_{s \rightarrow \infty} \frac{1}{s} p \left( \left. \frac{R}{s} \right
| \tau \right ) = \frac{1}{\sigma(\tau)} \exp{- \frac{R}{\sigma(\tau)}}
\end{equation}

\noindent In the \emph{orthogonal case} $(\beta=1)$, for example
\begin{equation} \label{32}
\sigma(\tau) = \left \{\begin{array}{ll} 4 - 2 \left | \tau  \right | + \left |
\tau \right | \ln{1 + 2\left | \tau  \right | }  & \quad \textrm{if} \,  \left |
\tau \right | \leq 1 \\
2 +    \left | \tau  \right | \ln{\frac{2 \left | \tau  \right | +1}{2\left |
\tau  \right | -1 }} & \quad \textrm{if}\,  \left | \tau  \right |   \geq 1
\end{array} \right.
\end{equation}

\noindent One can notice the \emph{singularity at the Heisenberg time} $\tau =1$
and the fact that $\sigma(\infty)$ exists.

\noindent However if we \emph{smooth out} in time $R(\tau)$, taking for example
\begin{equation} \label{33}
\overline{R} = \frac{1}{\tau_1- \tau_0} \int_{\tau_0}^{\tau_1} \, R(\tau) d\tau
\end{equation}
then we get a \emph{selft-averaging quantity}
\begin{equation} \label{34}
\lim_{s \rightarrow \infty} \frac{1}{s} p \left ( \frac{\overline{R}}{s} \right)
= \delta \left ( \overline{R}- \overline{\sigma} \right)
\end{equation}
with
\begin{equation} \label{35}
\sigma = \frac{1}{\tau_1 - \tau_0} \int_{\tau_0}^{\tau_1} d\tau \, \sigma(\tau)
\end{equation}

\noindent Some numerical work on chaotic billiards [5], in the large window limit, confirm
this exponential distribution. Integrable billiards show a very different
behaviour [5].

Finally, we would like to mention the fact that Wigner's energy level statistics
can be obtained for models, where eigenvalues and eigenvectors are correlated.
We think therefore that the study of quantum dynamics could discriminate between
such models and those we have considered where they are uncorrelated.
\section*{Acknowledgements}
This work is dedicated to F.Y. Wu for his $70^{th}$ birthday.

\end{document}